\documentclass[5p]{elsarticle}
\usepackage[english]{babel}
\usepackage[utf8]{inputenc}
\usepackage{hyperref}
\usepackage{amssymb}
\usepackage{amsmath}
\usepackage{graphicx}
\usepackage{color}
\usepackage{natbib}

\newcommand{\lign}[1]{\begin{align} #1 \end{align}}
\newcommand{\Tr}[1]{\mathrm{Tr}[#1]}
\newcommand{\incases}[1]{\begin{cases} #1 \end{cases}}
\newcommand{\lrangle}[1]{\langle #1 \rangle}
\newcommand{\pbracket}[1]{\left( #1 \right)}

\newcommand{\pderiv}[2]{\frac{\partial #1}{\partial #2}}

\begin{document}
\date{\today}
\author{Christian Holm Christensen}
\address{Niels Bohr International Academy and Discovery Center, Niels Bohr Institute, University of Copenhagen, Blegdamsvej 17, DK-2100 Copenhagen, Denmark}
\title{Exact Large-$N_{c}$ Solution of an Effective Theory for Polyakov Loops at Finite Chemical Potential}

\begin{abstract}
Using the strong coupling expansion we calculate the analytically exact order parameter for the effective theory of Polyakov loops with applied chemical potential in the large-$N_{c}$ limit.
\end{abstract}

\maketitle

\section{Introduction}
In recent years the order parameters of Quantum Chromodynamics at finite temperature and chemical potential ($\mu$) have gained renewed attention. At least part of this comes from the need to know where and when phase transitions occur in quark matter experiments, such as the observations of quark-gluon plasma at RHIC and LHC-ALICE.

However, due to the confining nature of QCD, the usual perturbative approach breaks down at low energies, meaning that other methods need to be employed for quantative results in this regime. A widely succesful method is the lattice formulation of QCD \cite{Wilson:1974sk} proposed by Wilson in 1974 in which the challenges lie in the evaluation of group integrals over $SU(3)$.

Aside from the significant challenges in the analytical evaluation of these integrals, numerical calculations also run into trouble in the form of the sign problem when the chemical potential is applied. As such it is instructive to work in the $N_{c} \to \infty$ limit, which, due to the $1/N_{c}$ expansion of QCD, provides a reasonably accurate approximation to $N_{c} = 3$ but allows for a simplification of some problems. Here we will present analytically exact calculations within the effective theory of Polyakov loops in the $N_{c} \to \infty$ limit and determine an expression for the order parameter of QCD in the case of an applied real quark chemical potential.

\section{Lattice Formulation}
Using the lattice formulation for a pure gauge theory proposed by Wilson \cite{Wilson:1974sk} the Yang-Mills action can be written as
\lign{
  S = \frac{\beta}{2N_{c}} \sum_{p} \Tr{U_{p} + U_{p}^{\dagger}}
}
where $\beta = \frac{2N_{c}}{g^{2}}$ and $U_{p} = U_{ij}U_{jk}U_{kl}U_{li}$ are the plaquette variables taking their values in the fundamental representation of $SU(N_{c})$. Quarks are introduced by using Wilson's method of heavy fermions by adding (in the notation by Creutz \cite{Creutz:1984mg})
\lign{
  S_{q} = \frac{1}{2}\kappa_{f}\sum_{\{i,j\}} \bar{\psi}_{i}(1+\gamma_{\mu}e_{\mu})U_{ij}\psi_{j} + \sum_{i}\bar{\psi}_{i}\psi_{i} 
}
where the $\{i,j\}$ sum is over all nearest-neighbour pairs with one term for each ordering of $i$ and $j$, $e_{\mu}$ is a unit vector from $i$ to $j$ and $\kappa_{f} = \frac{1}{aM_{f}+1+D}$ is the flavour-dependent hopping parameter with $D$ being the number of spatial dimensions. At this point it is straightforward to integrate out the fermionic degrees of freedom and expand the fermion determinant in terms of $\kappa_{f}$
\lign{
  \det(1-\kappa_{f}M(U)) = \exp\pbracket{-\sum_{L=1}^{\infty}\frac{\kappa_{f}^{L}}{L}\Tr{M^{L}(U)}}
}
where the sum is over all loops without backtracking. A chemical potential is added by multiplying the positively (negatively) oriented time-like products by $e^{a\mu}$ ($e^{-a\mu}$) in likeness of a fourth component of the gauge field \cite{Hasenfratz:1983ba}.
At this point the effective theory of Polyakov loops can be derived using the strong coupling expansion together with the option of gauge fixing any maximal tree of link variables $U_{ij} = 1$ \cite{Bartholomew:1983jv} or by using a Kadanoff-inspired bond-moving procedure \cite{Ogilvie:1983ss}. To leading order in both $\beta$ and $\kappa$ the theory takes the form
\lign{
  S_{eff}(J,&he^{aN_{\tau}\mu},he^{aN_{\tau}\mu}) \nonumber\\
  = \frac{1}{2}J &\sum_{x,j} [W(x)W^{\dagger}(x+j) + W^{\dagger}(x)W(x+j)] \nonumber\\
  + N_{c}&\sum_{x}[he^{aN_{\tau}\mu}W(x) + he^{-aN_{\tau}\mu}W^{\dagger}(x)]\label{eq:PolyAct}
}
where $W(x) = \mathrm{Tr}\prod\limits_{n=0}^{N_{\tau}-1} U_{0i}$ is the thermal Polyakov loop, $j>0$ runs over the nearest neighbours only and $J = 2(\beta/2N^{2}_{c})^{N_{\tau}}$ and $h=2\frac{N_{f}}{N_{c}}\kappa^{N_{\tau}}$ to leading order for $N_{f}$ mass degenerate quark flavours. Here $N_{\tau}$ is the number of lattice sites in the temporal direction. The values of $J$ and $h$ get renormalized by taking into account the spatial link integrations. For notational simplicity we define $\hat{\mu} \equiv aN_{\tau}\mu$ from now on.
The purpose of this paper is to derive the analytically exact expectation value, $\lrangle{W}$, of the Polyakov loop in the $N_{c} \to \infty$ limit, keeping the ratio $\frac{N_{f}}{N_{c}}$ fixed.
\section{The Expectation Value}
The discussion below is largely analoguous to that of \cite{Damgaard:1986mx} but with a few crucial exceptions in order to treat the case of an applied chemical potential. Starting from the action (\ref{eq:PolyAct}) and writing
\lign{
  S&(J,0,0) = \frac{1}{2}J\sum_{x,j}\left\{\lrangle{W(x)}W^{\dagger}(x+j)\right. \nonumber \\
  &+W(x)\lrangle{W^{\dagger}(x+j)}-\lrangle{W(x)}\lrangle{W^{\dagger}(x+j)} \\
  &\left.+[W(x)-\lrangle{W(x)}][W^{\dagger}(x+j)-\lrangle{W^{\dagger}(x+j)}] + h.c \right\} \nonumber
}
the last term does not contribute in the $N_{c} \to \infty$ limit due to factorization. Translational invariance allows the $j$-sum to be performed for arbitrary number of spatial dimensions
\lign{
   S(J,he^{\hat{\mu}},&he^{-\hat{\mu}}) = N_{c}\sum_{x}[he^{\hat{\mu}}W(x)+he^{-\hat{\mu}}W^{\dagger}(x)] \\ + &JD \sum_{x} [\lrangle{W}W^{\dagger}(x) + \lrangle{W^{\dagger}}W(x)-\lrangle{W}\lrangle{W^{\dagger}}] \nonumber
}
Notice here that $\lrangle{W}^{\dagger} \neq \lrangle{W^{\dagger}}$ since the action is non-Hermitian due to the chemical potential being real. In fact $\lrangle{W}^{\dagger} = \lrangle{W}$ and $\lrangle{W^{\dagger}}^{\dagger}=\lrangle{W^{\dagger}}$ due to the Haar measure, meaning both expectations are real. The action can also be expressed as
\lign{
  S(J,he^{\hat{\mu}},he^{-\hat{\mu}}) &= S(0,h_{+},h_{-}) - JD \sum_{x} \lrangle{W}\lrangle{W^{\dagger}}
}
with $h_{\pm}$ given by the self-consistency relations:
\lign{
  h_{+} &= \frac{JD}{N_{c}}\lrangle{W^{\dagger}} + he^{\hat{\mu}} \\
  h_{-} &= \frac{JD}{N_{c}}\lrangle{W} + he^{-\hat{\mu}}
}
As usual, the expectation values are given by
\lign{
  \lrangle{W} &= \frac{1}{Z} \int\limits_{SU(N)}dW \pbracket{W \exp[S(0,h_{+},h_{-})]} \\
  \lrangle{W^{\dagger}} &= \frac{1}{Z} \int\limits_{SU(N)}dW \pbracket{W^{\dagger} \exp[S(0,h_{+},h_{-})]}
}
with
\lign{
  Z(J,&h_{+},h_{-}) = N(J)\int\limits_{SU(N)} dW \exp[S(0,h_{+},h_{-})] \\
  &= N(J)\int\limits_{SU(N)} dW \exp\left[N_{c}\sum_{x}[h_{+}W+h_{-}W^{\dagger}]\right] \nonumber
}
where $N(J) = \exp(-JD\sum_{x} \lrangle{W}\lrangle{W^{\dagger}})$ is a J-dependent prefactor.
\section{SU($N_{c}$) to U($N_{c}$)}
\begin{figure}[htb]
  \centering
  \includegraphics[trim = 10mm 0mm 10mm 0mm, width=0.4\textwidth]{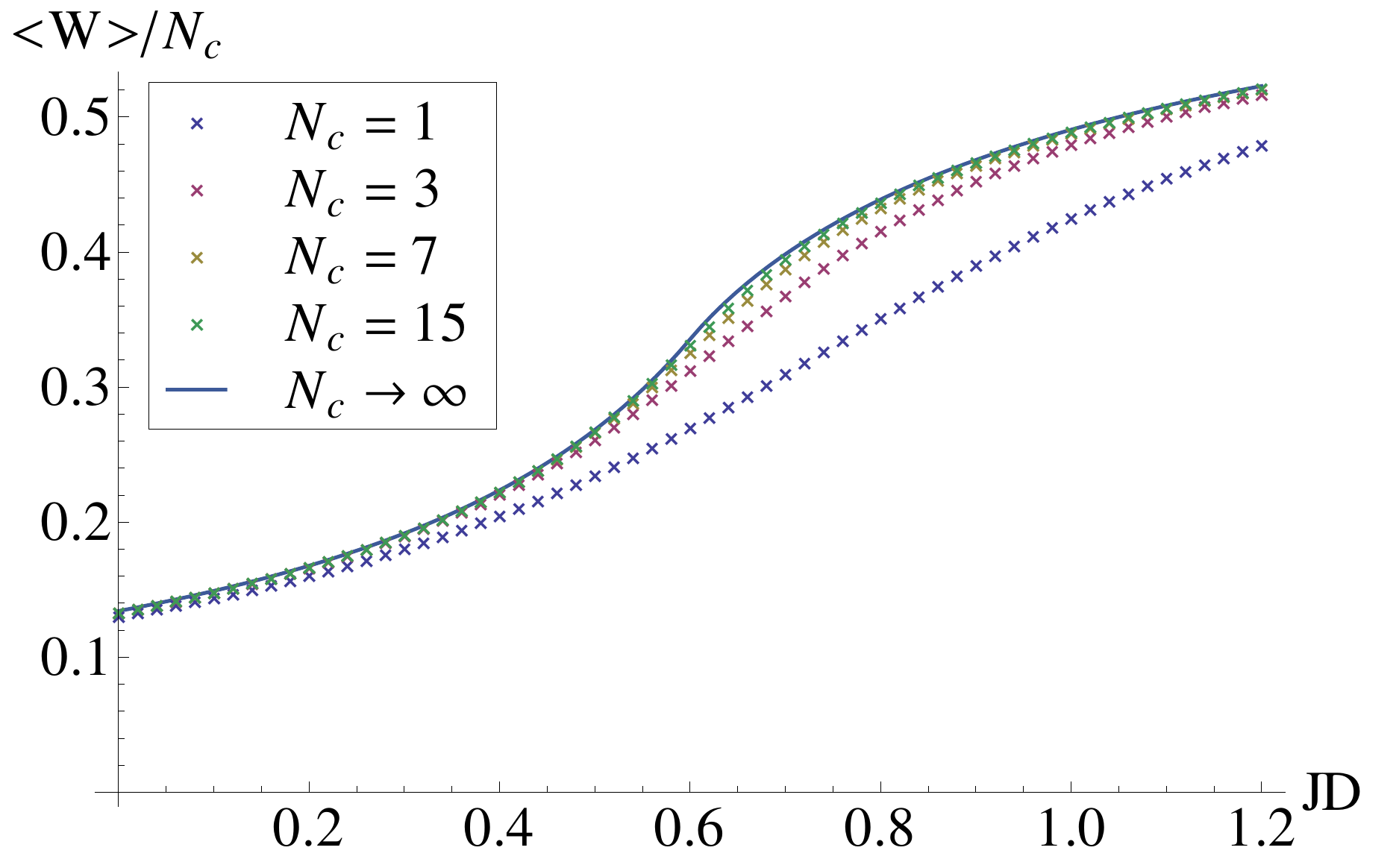}
  \caption{Plot of $\lrangle{W}/N_{c}$ for $h = 0.2$ and $\hat{\mu} = 0.4$ for $N_{c} = 1,3,7,15$ and $\infty$ in the case of $U(N_{c})$.}
  \label{fig:ABeff}
\end{figure}

\begin{figure}
  \centering
  \includegraphics[trim = 5mm 5mm 5mm 5mm, width=0.4\textwidth]{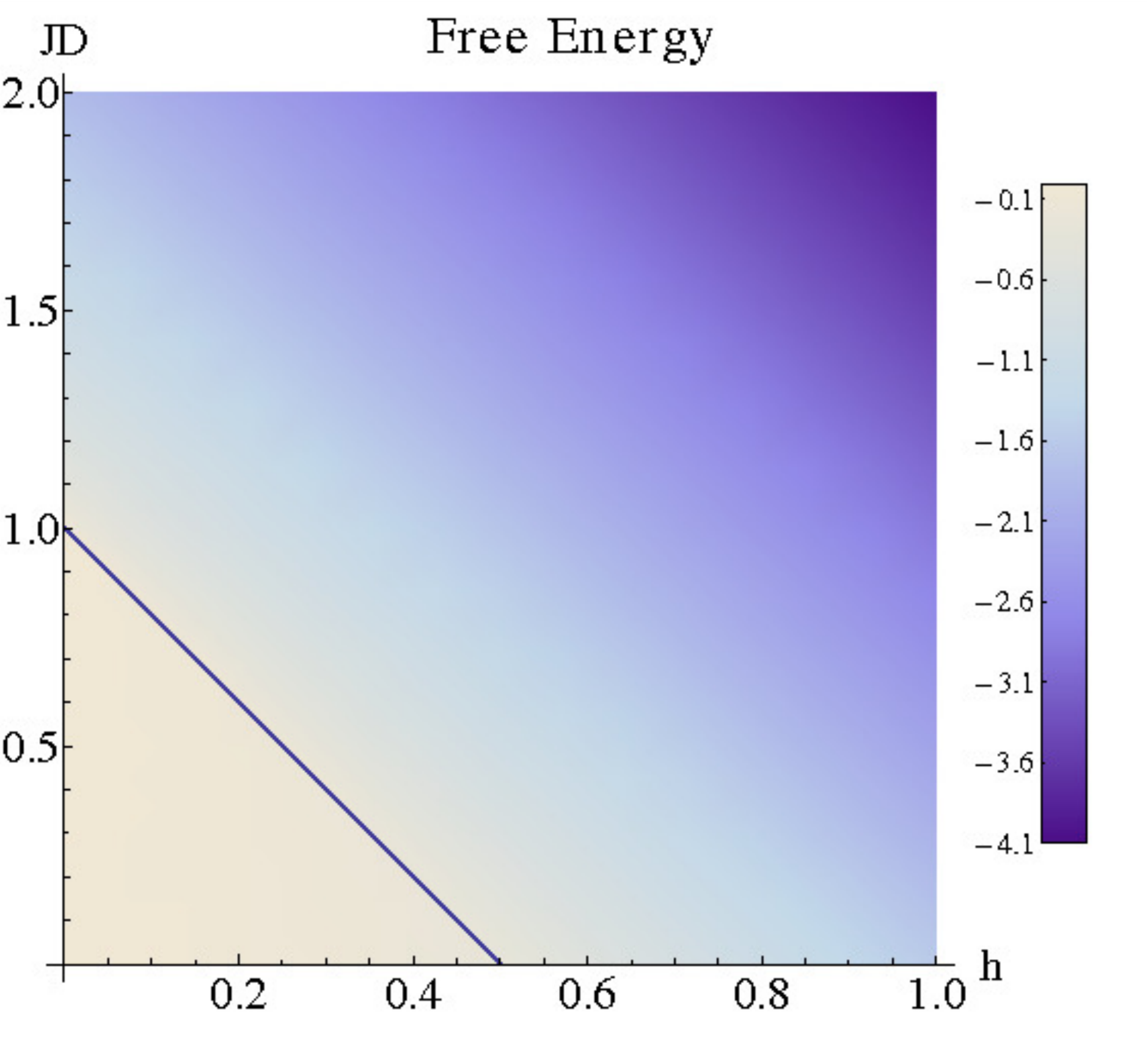}
  \caption{Plot of the free energy F(JD,h) with the transition line at $JD = 1-2h$ added.}
  \label{fig:FreeEn}
\end{figure}

In the large-$N_{c}$ limit SU($N_{c}$) integration is equivalent to U($N_{c}$) integration. The relevant integral, as calculated by \cite{Schlittgen:2002tj}, is
\lign{
  I&(A,B) = \int\limits_{U(N)} dU e^{\frac{1}{2}\Tr{AU+BU^{\dagger}}} \nonumber \\
  &= 2^{\frac{N(N-1)}{2}}\left[\prod_{n=1}^{N-1} n! \right] \frac{\det[\lambda_i^{j-1}I_{j-1}(\lambda_i)]}{\Delta(\lambda^2)}
  \label{eq:GenUInt}
}
where $\lambda_{i}^{2}$ is the eigenvalues of $AB$ and $\Delta(x)$ is the Vandermonde determinant
\lign{
  \Delta(x) = \prod_{i<j}^{N} (x_{j} - x_{i})
}
In the relevant case $A = h_{+}I$ and $B = h_{-}I$ where $I$ is an $N_{c} \times N_{c}$ identity matrix which means that both nominator and denominator of eq. (\ref{eq:GenUInt}) have $\frac{1}{2}N(N-1)$ zeros. Removing these using L'Hôpital's rule reduces the integral to
\lign{
  I(h_{+},h_{-},0) &= \int\limits_{U(N_{c})}dW \exp[N_{c}h_{+}W+N_{c}h_{-}W^{\dagger}] \\
  &= \det[I_{j-i}(2N_{c}\sqrt{h_{+}h_{-}})] \label{eq:UnitInt}
}
up to an overall, and in our case irrelevant, normalization constant. Amazingly enough this equation means that the integral, $I$, only depends on the product of $h_{+}$ and $h_{-}$, simplifying the problem of calculating the expectation values markedly. Reexpressing the partition function:
\lign{
  Z(J,h_{+},h_{-}) = Z(J,\sqrt{h_{+}h_{-}},\sqrt{h_{+}h_{-}})
  \label{eq:PartToProd}
}
means that the expectation values can be expressed as
\lign{
  \lrangle{W} &= \frac{1}{N_{c}N_{x}Z}\pderiv{}{h_{+}}Z(J,\sqrt{h_{+}h_{-}},\sqrt{h_{+}h_{-}}) \nonumber \\
  &= \frac{1}{2N_{c}N_{x}Z}\sqrt{\frac{h_{-}}{h_{+}}}\pderiv{}{g}Z(J,g,g) \\
  \lrangle{W^{\dagger}} &= \frac{1}{N_{c}N_{x}Z}\pderiv{}{h_{-}}Z(J,\sqrt{h_{+}h_{-}},\sqrt{h_{+}h_{-}}) \nonumber \\
  &= \frac{1}{2N_{c}N_{x}Z}\sqrt{\frac{h_{+}}{h_{-}}}\pderiv{}{g}Z(J,g,g)
}
for $g = \sqrt{h_{-}h_{+}}$, $h_{+},h_{-} \neq 0$ and $N_{x} = \sum\limits_{x}$. The case of $h_{+} = 0$ and $h_{-} = 0$ is trivially solved and yields $\lrangle{W} = \lrangle{W^{\dagger}} = 0$. Notice that from these expressions $h_{+}\lrangle{W}=h_{-}\lrangle{W^{\dagger}}$, which, by using the self-consistency relations, also leads to $h_{+}he^{-\hat{\mu}} = h_{-}he^{\hat{\mu}}$. Using \cite{Damgaard:1986mx,Gross:1980he} the integration can be performed:
\lign{
  \frac{h_{+}\lrangle{W}}{N_{c}} = \frac{h_{-}\lrangle{W^{\dagger}}}{N_{c}} = \incases{\qquad h_{+}h_{-} & \textrm{for  } \sqrt{h_{+}h_{-}} < \frac{1}{2}\\
    \sqrt{h_{+}h_{-}}-\frac{1}{4} & \textrm{for  } \sqrt{h_{+}h_{-}} \geq \frac{1}{2}}
  \label{eq:NInfInt}
}
Solving the self-consistency equations in the case of $\sqrt{h_{+}h_{-}}$\\$< \frac{1}{2}$ is straightforward, but in the case of $\sqrt{h_{+}h_{-}} \geq \frac{1}{2}$ there is a small detail which requires mentioning. Rewriting the integration result (\ref{eq:NInfInt})
\lign{
 h_{+}\pbracket{\lrangle{W}-\sqrt{\frac{h_{-}}{h_{+}}}} = h_{-}\pbracket{\lrangle{W^{\dagger}}-\sqrt{\frac{h_{+}}{h_{-}}}} = -\frac{1}{4}
}
we can use that for $h \neq 0$ then $\frac{h_{+}}{h_{-}} = e^{2\hat{\mu}}$ to simplify the expression. To make this step cover the case of $h = 0$ requires setting $\hat{\mu} = 0$ simultaneously, such that $\frac{h_{+}}{h_{-}} = 1$ which is the solution when $h=0$ from the beginning. Now, having taken the preliminary steps, the rest of the derivation is simple and leads to the solutions
\lign{
  \frac{\lrangle{W}}{N_{c}} &= \incases{\frac{h}{1-JD}\exp[-\hat{\mu}] \\\frac{1}{2} \pbracket{1 - \frac{h}{JD} + \sqrt{\pbracket{1 + \frac{h}{JD}}^{2} - \frac{1}{JD}}}\exp[-\hat{\mu}] }
}
and
\lign{
  \frac{\lrangle{W^{\dagger}}}{N_{c}} &= \incases{\frac{h}{1-JD}\exp[\hat{\mu}] \\\frac{1}{2} \pbracket{1 - \frac{h}{JD} + \sqrt{\pbracket{1 + \frac{h}{JD}}^{2} - \frac{1}{JD}}}\exp[\hat{\mu}] }
}
with a transition point between the two solutions at $JD = 1 - 2h$. As can be seen from the solutions, simply setting $h = 0$ does not remove the $\mu$-dependence; it is necessary to also set $\mu = 0$, revealing that $h=0$ cannot be reached in the $h \to 0$ limit of the solution. Figure \ref{fig:ABeff} showing the convergence of the numerically calculated $\lrangle{W}$ values for $U(N_{c})$ when subjected to the self-consistency relations in the case of $N_{c} = 1,3,7,15$ together with the analytical expression in the $N_{c} \to \infty$ limit.
The associated free energy is found to be
\lign{
  F(J) = \incases{\qquad \qquad \quad JD \lrangle{W}\lrangle{W^{\dagger}} - h_{+}h_{-} \\ JD \lrangle{W}\lrangle{W^{\dagger}} - 2\sqrt{h_{+}h_{-}} + \frac{1}{2}\ln(2\sqrt{h_{+}h_{-}}) + \frac{3}{4}}
}
with a third order phase-transition point at $JD = 1 - 2h$ in which $\lrangle{W} = \frac{1}{2}\exp(\hat{\mu})$ and $\lrangle{W^{\dagger}} = \frac{1}{2}\exp(-\hat{\mu})$. The transition line is illustrated in fig. \ref{fig:FreeEn}. Inserting the solutions the $JD < 1 - 2h$ part of the free energy can be simplified and written as
\lign{
    F(J) = -\frac{h^{2}}{1-JD} \quad \textrm{for } JD < 1 - 2h
}
The free energy, like the partition function in \ref{eq:PartToProd}, is independent of the chemical potential, since the factors of $e^{\hat{\mu}}$ and $e^{-\hat{\mu}}$ cancel - a consequence of this is that the quark number density is zero
\lign{
  n_{q} = \lrangle{\psi^{\dagger}\psi} = \pderiv{}{\mu}F = 0
}
This result is an artifact of $U(N_{c})$ and not $SU(N_{c})$ - a claim which is supported in a recent mean field study \cite{Greensite:2012xv} further investigating the $U(N_{c})$ and $SU(N_{c})$ difference. Note that, surprisingly, this means that the $N_{c} \to \infty$ limit of the $U(N_{c})$ and $SU(N_{c})$ theories \emph{differ} even though the difference between the two groups vanishes as $1/N_{c}^{2}$.

For recent numerical work for $SU(3)$ in the case of a real chemical potential, we refer the reader to \cite{Aarts:2011zn, Mercado:2012ue, Gattringer:2011gq} where complex Langevin techniques and the flux representation is used. Work on a related model incorperating higher order contributions to the effective theory is found in \cite{Langelage:2010yr} and \cite{Fromm:2011qi} respectively. It is remarkable that the phase diagram of the $N_{c}=3$ theory \cite{Fromm:2011qi} looks qualitively similar to that of fig. \ref{fig:FreeEn}.

\section{Conclusion}
We have found an analytically exact expression for the order parameter of QCD, $\lrangle{W}$, within the effective theory of Polyakov loops with an applied chemical potential in the limit $N_{c} \to \infty$. Furthermore the free energy has been calculated and displayed a third order phase transition at $JD = 1 - 2h$. The free energy, and hence also the point of the phase transition, was found to be $\hat{\mu}$ independent, in agreement with large-$N_{c}$ counting. The expectation values of the Polyakov loops display the dependence on quark chemical potential that could be expected from intuitive arguments.

\section{Acknowledgements}
The author would like to thank Poul H. Damgaard and Kim Splittorf for helpful insights and discussions.


\vspace{0.4cm}
\begin{thebibliography}{99}
\bibitem{Wilson:1974sk}
  K.~G.~Wilson,
  Phys.\ Rev.\  {\bf D10}, 2445-2459 (1974).

\bibitem{Creutz:1984mg} 
  M.~Creutz,
  Cambridge, Uk: Univ. Pr. ( 1983) 169 P. (Cambridge Monographs On Mathematical Physics)

\bibitem{Hasenfratz:1983ba}
  P.~Hasenfratz and F.~Karsch,
  Phys.\ Lett.\  B {\bf 125}, 308 (1983).

\bibitem{Bartholomew:1983jv}
  J.~Bartholomew, D.~Hochberg, P.~H.~Damgaard and M.~Gross,
  Phys.\ Lett.\  B {\bf 133}, 218 (1983).

\bibitem{Ogilvie:1983ss}
  M.~Ogilvie,
  Phys.\ Rev.\ Lett.\  {\bf 52}, 1369 (1984).

\bibitem{Damgaard:1986mx}
  P.~H.~Damgaard and A.~Patkós,
  Phys.\ Lett.\  B {\bf 172}, 369 (1986)

\bibitem{Schlittgen:2002tj}
  B.~Schlittgen and T.~Wettig,
  J.\ Phys.\  A {\bf A36}, 3195 (2003)
  [arXiv:math-ph/0209030]

\bibitem{Gross:1980he}
  D.~J.~Gross and E.~Witten,
  Phys.\ Rev.\  D {\bf 21}, 446 (1980).

\bibitem{Greensite:2012xv} 
  J.~Greensite and K.~Splittorff,
  arXiv:1206.1159 [hep-lat].

\bibitem{Aarts:2011zn} 
  G.~Aarts and F.~A.~James,
  JHEP {\bf 1201}, 118 (2012)
  [arXiv:1112.4655 [hep-lat]].

\bibitem{Mercado:2012ue} 
  Y.~D.~Mercado and C.~Gattringer,
  Nucl.\ Phys.\ B {\bf 862}, 737 (2012)
  [arXiv:1204.6074 [hep-lat]].

\bibitem{Gattringer:2011gq} 
  C.~Gattringer,
  Nucl.\ Phys.\ B {\bf 850}, 242 (2011)
  [arXiv:1104.2503 [hep-lat]].

\bibitem{Langelage:2010yr}
  J.~Langelage, S.~Lottini and O.~Philipsen,
  JHEP {\bf 1102} (2011) 057
  [Erratum-ibid.\  {\bf 1107} (2011) 014]
  [arXiv:1010.0951 [hep-lat]].
  
\bibitem{Fromm:2011qi}
  M.~Fromm, J.~Langelage, S.~Lottini and O.~Philipsen,
  JHEP {\bf 1201} (2012) 042
  [arXiv:1111.4953 [hep-lat]].

\end{thebibliography}
\end{document}